\documentclass{emulateapj}
\submitted{Accepted to ApJL}

\newcommand{\rcore}{r_{\rm core}}
\newcommand{\mtot}{M_{\rm tot}}
\newcommand{\qp}{Q_{\rm p}}
\newcommand{\qunit}{\textrm{M}_\odot \textrm{pc}^{-3} (\textrm{km}/\textrm{s})^{-3}}
\newcommand{\vmax}{V_{\rm max}}
\newcommand{\rhocore}{\rho_0}

\shorttitle{The Case Against Warm Dark Matter}
\shortauthors{Kuzio de Naray et al.}

\begin{document}

\title{The Case Against Warm or Self-Interacting Dark Matter as 
  Explanations for Cores in Low Surface Brightness Galaxies}

\author{Rachel Kuzio de Naray\altaffilmark{1}, Gregory D. Martinez,
  James S. Bullock, Manoj Kaplinghat} 

\affil{Center for Cosmology, Department of Physics and Astronomy,
University of California, Irvine, CA 92697-4575}

\altaffiltext{1}{NSF Astronomy and Astrophysics Postdoctoral Fellow}

\email{kuzio@uci.edu}
\email{gmartine@uci.edu}
\email{bullock@uci.edu}
\email{mkapling@uci.edu}

\begin{abstract}
Warm dark matter (WDM) and self-interacting dark matter (SIDM) are
often motivated by the inferred cores in the dark matter halos of low
surface brightness (LSB) galaxies.  We test thermal WDM, non-thermal
WDM, and SIDM using high-resolution rotation curves of nine LSB
galaxies.  We fit these dark matter models to the data and determine
the halo core radii and central densities.  While the minimum core
size in WDM models is predicted to decrease with halo mass, we find
that the inferred core radii increase with halo mass and also cannot
be explained with a single value of the primordial phase space
density.  Moreover, if the core size is set by WDM particle
properties, then even the smallest cores we infer would require
primordial phase space density values that are orders of magnitude
smaller than lower limits obtained from the Lyman alpha forest power
spectra.  We also find that the dark matter halo core densities vary
by a factor of about 30 from system to system while showing no systematic
trend with the maximum rotation velocity of the galaxy. This strongly
argues against the core size being directly set by large
self-interactions (scattering or annihilation) of dark matter.  We
therefore conclude that the inferred cores do not provide motivation
to prefer WDM or SIDM over other dark matter models.
\end{abstract}


\section{Introduction}

In the prevailing theory of galaxy formation, galaxies assemble inside
cuspy cold dark matter (CDM) halos.  Dissipationless CDM simulations
show that these halos have steeply rising central densities that roll
over from $\rho \sim r^{-1}$ at $\sim$\,1 kpc scales to $\rho \sim
r^{-0.8}$ at $\sim$\,100 pc scales
\citep[e.g.,][]{Navarro04,Graham06,Navarro08}.  There has been
continued debate in the literature over this prediction as numerous
observational results indicate that rotation curve data are often more
consistent with dark matter halos having a roughly constant density
core
\citep[e.g.,][]{Flores,Moore94,MRdB,Marchesini,Gentile05,Simon05,Kuzio06,Kuzio08,dB08,dB09}.

The mismatch between the dissipationless CDM simulations and galaxy
rotation curve data has motivated a serious exploration of warm dark
matter (WDM) models
\citep{Hogan,Avila01,Abazajian01,kaplinghat05,cembranos05,strigari07} 
and self-interacting dark matter (SIDM) \citep{Spergel00,Kaplinghat00}
as alternatives to CDM.  The goal is to maintain the success of CDM on
large scales while producing cores in the dark matter distribution of
small halos.

Two generic classes of WDM particles are thermal particles that are
light and which kinetically decouple when relativistic
\citep{Blumenthal,dw94,Asaka06}, and particles that are as massive as typical 
CDM particles but populated by decays in the early universe
\citep{kaplinghat05,cembranos05}.  In both classes there are
regions of parameter space that are endowed with large particle
velocities and correspondingly low primordial phase space densities
$\qp = \bar{\rho}/\bar{\sigma}^3$. The initial phase space density
imposes a limit on the central phase space density of collapsed WDM
halos for both thermal WDM particles \citep{TG79} and non-thermal
particles arising from early-decays \citep{kaplinghat05}. Warmer dark
matter models have lower $\qp$ and correspondingly larger limiting
core sizes at fixed halo mass.  If the cores in dark matter halos are
set by primordial phase-space constraints, then for a given $\qp$
there are predicted relationships between halo structural parameters
(e.g., core radius and central density).  This provides a means to
evaluate WDM as a solution to the cusp-core problem.

In SIDM models, the interactions lead to either thermalization or
particle loss and the subsequent formation of a core.  Models that
have been studied are those with large cross sections for scattering
\citep{Spergel00,Firmani00} or annihilations
\citep{Kaplinghat00}. In the  
simplest case where the s-wave contributions dominate the cross section,
we expect all core densities to cluster around a common value. For
more complicated models, the core density should correlate with the velocity
dispersion in the core.

In this Letter we fit high-resolution rotation curves of low surface
brightness (LSB) disk galaxies using cored dark matter density
profiles of the type expected in WDM and SIDM models. These galaxies
are dark matter-dominated down to small radii \citep[e.g.,][but see
Fuchs 2003]{deBlok96}, therefore providing a good laboratory for
testing dark matter halo predictions.  In Section 2, we describe the
density profiles we use for models of thermal WDM and non-thermal WDM
from early decays.  In Section 3, we fit the galaxy data with these
halo models and determine the sizes of the halo cores.  In Section 4,
we discuss predictions for particle dark matter models based on the
measured core sizes.  A summary is presented in Section 5.

\section{Cored Dark Matter Halo Models}

Provided below are brief descriptions of the cored halo profiles used
in our analysis. We consider four different profiles to bracket both a
range of behavior in the outer region of the density profile and the
rapidity of the transition from the core to the outer region. In all
cases, the profiles are set by two independent parameters: a scale
density and scale radius. Recent WDM \citep{colin08} and SIDM
\citep{Dave01} simulations support this assumption.  The specific
forms of the profiles we use are motivated by arguments in the limit
of no phase space mixing. However, given the wide range of behaviors
encapsulated by these profiles, we also use them for our analysis of
SIDM models.

\subsection{Thermal Warm Dark Matter}

The density profile for the thermal WDM halo, in the limit of no phase
space mixing, is subject to the constraint that as r $\rightarrow$ 0,
$\rho \rightarrow \rhocore[1 - (r/\rcore)^{2}]$.  This ensures that
the phase space density is finite for all values of total energy. This
is necessary because the phase space density of thermal particles in
the early universe is finite for all momenta. 
We assume the following form for the thermal WDM halo density profile    
\begin{equation}
\rho_{\rm{Th}}(r) =
\frac{\rhocore}{\left[1+{1\over \alpha}\left(\frac{r}{\rcore}\right)^{2}\right]^{\alpha}}, 
\end{equation}
where $\rhocore$ is the central core density and $\rcore$ is the core
radius.

We test the thermal WDM model with $\alpha=1$ and $\alpha=2$.  The
$\alpha=1$ case can be recognized as the cored pseudoisothermal halo
traditionally used in rotation curve fitting.  We use this case to
test how sensitive our results are to the assumed density profile.

The $\alpha=1$ (Th1) and $\alpha=2$ (Th2) rotation curves are
\begin{eqnarray}
V^2_{\rm{Th1}}(r) &=& 4\pi G\rhocore \rcore^{2}\left[1 -
\frac{1}{x}\arctan\left(x\right)\right],\nonumber\\
V^2_{\rm{Th2}}(r) &=& 4\pi G\rhocore 
\rcore^{2}\left[\frac{\arctan(x)}{x} - \frac{1}{x^{2}+1}\right],
\end{eqnarray}
with $x = r/(\sqrt\alpha \rcore)$.

\subsection{Non-thermal Warm Dark Matter from Early Decays}

For particles populated by decays of massive particles in the early
universe, the density profile is subject to the constraint that as r
$\rightarrow$ 0, $\rho \rightarrow \rhocore(1 - r/\rcore)$ in the
limit of no phase space mixing. This is a milder core than that
created by thermal WDM particles.  We require that the density profile
be of the form 
\begin{equation}
\rho_{\rm{ED}}(r) =
\frac{\rhocore}{\left(1+\frac{r}{\alpha \rcore}\right)^{\alpha}},
\end{equation}
where $\rhocore$ is the central core density and $\rcore$ is the core
radius.  For further discussion about the form of these profiles, we
refer the reader to \citet{Martinez}.

We test the early decay model with $\alpha = 3$ and $\alpha = 4$.  The
choice of these exponents is motivated by the range of slopes seen in
CDM simulations in the outer regions of the halos
\citep{Navarro04,Graham06,Navarro08}. Because the outer regions are
built up from accreted dark matter particles, there should be no
difference between cold and warm dark matter predictions
\citep{colin08}.

The rotation curves corresponding to the $\alpha=3$ (ED3) and
$\alpha=4$ (ED4) cases  are
\begin{eqnarray}
V^2_{\rm{ED3}}(r) &=& 
36\pi G \rhocore \rcore^{2}
\left[\frac{\ln(1+x)}{x} - \frac{2+3x}{2(1+x)^{2}} \right],\nonumber\\
V^2_{\rm{ED4}}(r) &=& 64\pi G \rhocore \rcore^{2}
\frac{x^2}{3(1+x)^3},
\label{eq:EDrotcurve}
\end{eqnarray}
with $x = r/(\alpha \rcore)$.

\begin{figure*}
\epsscale{1.0} \plotone{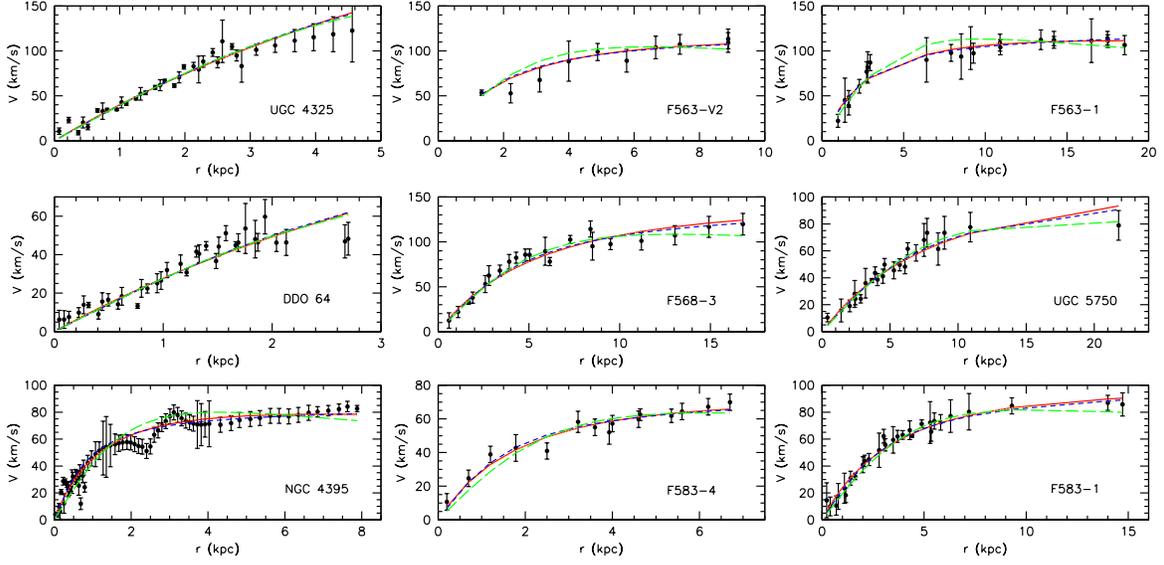}
\caption{Observed LSB galaxy rotation curves with the best-fitting
early decay dark matter ($\alpha=3$: solid red, $\alpha=4$: dotted
orange) and  thermal WDM ($\alpha=1$: short-dash blue; $\alpha=2$:
long-dash green) halo fits overlaid. (A color version of this figure
is available in the online journal.)}
\end{figure*}

\begin{deluxetable*}{lccccccccccc}
\tabletypesize{\small}
\tablecaption{Best-Fit Cored Halo Parameters and Primordial Phase Space Densities} 
\tablewidth{0pt} 
\tablehead{
\colhead{} &\colhead{}  &\multicolumn{4}{c}{Early Decay DM $\alpha=3$} &\colhead{} &\multicolumn{5}{c}{Early Decay DM $\alpha=4$}\\
\cline{3-6} \cline{8-12} 
\colhead{Galaxy} &\colhead{}&\colhead{$\rcore$} &\colhead{$\rhocore$} &\colhead{$\chi^{2}_{r}$} &\colhead{$\qp$} &\colhead{} &\colhead{$\rcore$} &\colhead{$\rhocore$} &\colhead{$\chi^{2}_{r}$} &\colhead{$\qp$} &\colhead{$\mtot$} }
\startdata 
UGC 4325 & &4.6$\tablenotemark{a}$ &106$\tablenotemark{b}$ &3.1 &\nodata  & &4.6$\tablenotemark{a}$ &106$\tablenotemark{b}$ &3.1 &\nodata &\nodata\\ 
F563-V2  & &1.1$\pm$0.1 &188$\pm$30 &0.65 &4.8   & &1.3$\pm$0.2  &167$\pm$25   &0.74 &5.9 &9.8\\ 
F563-1   & &1.4$\pm$0.1 &106$\pm$16 &0.50 &3.2  & &1.8$\pm$0.1  &90$\pm$12    &0.50 &3.1 &14\\ 
DDO 64   & &2.7$\tablenotemark{a}$&57$\tablenotemark{b}$      &3.3  &\nodata   & &2.7$\tablenotemark{a}$    &57$\tablenotemark{b}$     &3.3  &\nodata &\nodata\\ 
F568-3   & &2.8$\pm$0.4 &40$\pm$6   &1.5  &0.71  & &3.0$\pm$0.4  &38$\pm$5     &1.4  &1.0 &28\\ 
UGC 5750 & &4.3$\pm$0.7 &11$\pm$1   &0.92 &0.42  & &4.8$\pm$0.8  &10$\pm$1     &0.89 &0.53 &30\\ 
NGC 4395 & &0.6$\pm$0.1 &346$\pm$42 &2.6  &21   & &0.7$\pm$0.1  &278$\pm$33   &3.0  &29 &2.6\\ 
F583-4   & &0.9$\pm$0.1 &98$\pm$22  &0.59 &13   & &1.1$\pm$0.2  &82$\pm$18    &0.68 &14 &2.9\\
F583-1   & &1.7$\pm$0.1 &48$\pm$4   &0.58 &2.9  & &2.1$\pm$0.2  &43$\pm$3     &0.54 &2.9 &11\\ 
\hline\\ 
& & \multicolumn{4}{c}{Thermal WDM $\alpha=1$}  & &\multicolumn{5}{c}{Thermal WDM $\alpha=2$}\\
\cline{3-6} \cline{8-12} 
Galaxy & &$\rcore$ &$\rhocore$ &$\chi^{2}_{r}$ &$\qp$ & &$\rcore$ &$\rhocore$ &$\chi^{2}_{r}$ &$\qp$ &$\mtot$\\ 
\hline\\ 
UGC 4325 & &4.1$\pm$1.0 &88$\pm$6     &3.1  &\nodata   & &4.3$\pm$1.0  &90$\pm$5    &3.1   &\nodata &\nodata\\ 
F563-V2  & &1.5$\pm$0.2 &118$\pm$18   &0.71 &72    & &2.5$\pm$0.2  &93$\pm$14   &1.4   &14 &4.3\\
F563-1   & &2.1$\pm$0.2 &66$\pm$9     &0.43 &35    & &3.8$\pm$0.2  &47$\pm$6    &0.69  &5.9 &7.3\\ 
DDO 64   & &2.7$\tablenotemark{a}$   &45$\tablenotemark{b}$     &3.1  &\nodata    & &2.7$\tablenotemark{a}$    &47$\tablenotemark{b}$    &3.1   &\nodata &\nodata\\ 
F568-3   & &3.8$\pm$0.4 &27$\pm$3     &1.2  &9.3   & &5.0$\pm$0.4  &25$\pm$2    &1.1   &3.4 &9.2\\ 
UGC 5750 & &5.7$\pm$0.8 &7.1$\pm$0.7  &0.83 &5.4   & &7.1$\pm$0.7  &7.0$\pm$0.2 &0.74  &2.4 &6.9\\ 
NGC 4395 & &0.7$\pm$0.1 &262$\pm$34   &2.9  &478   & &1.7$\pm$0.1  &121$\pm$16  &5.1   &42 &1.7\\
F583-4   & &1.3$\pm$0.2 &66$\pm$16    &0.67 &149    & &2.4$\pm$0.3  &38$\pm$8    &1.2   &26 &1.5\\ 
F583-1   & &2.5$\pm$0.2 &30$\pm$2     &0.50 &31    & &4.0$\pm$0.2  &22$\pm$1    &0.77  &7.0 &4.2\\ 
\enddata
\tablecomments{Best-fit halo parameters ($\rcore$, $\rhocore$), lower limits on the primordial phase space densities ($\qp$), and the total mass of the system, $\mtot$.  The units for $\rcore$
are kpc and the units for $\rhocore$ are 10$^{-3}$ M$_{\sun}$
pc$^{-3}$.  The units for $\qp$ are 10$^{-9}$ M$_\sun$ pc$^{-3}$ (km
s$^{-1}$)$^{-3}$.  The units for $\mtot$ are 10$^{10}$M$_{\sun}$. }
\tablenotetext{a}{upper limit}
\tablenotetext{b}{\,lower limit}
\end{deluxetable*}

\begin{figure}
\epsscale{1.0} 
\plotone{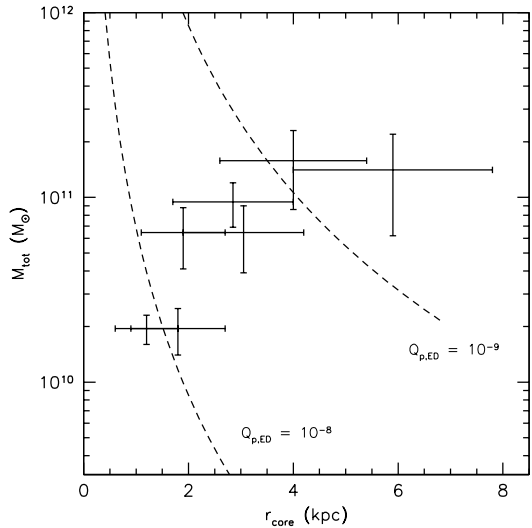}
\caption{Mass of each galaxy-halo system as a function of measured
halo core radius for the Early Decay $\alpha=4$ and Thermal WDM
$\alpha=2$ models.  Lines of constant $\qp$ for the early decay case
are overplotted.  Lines of constant $\qp$ for the thermal case are
similar in shape.}
\end{figure}

\begin{figure}
\epsscale{1.0} 
\plotone{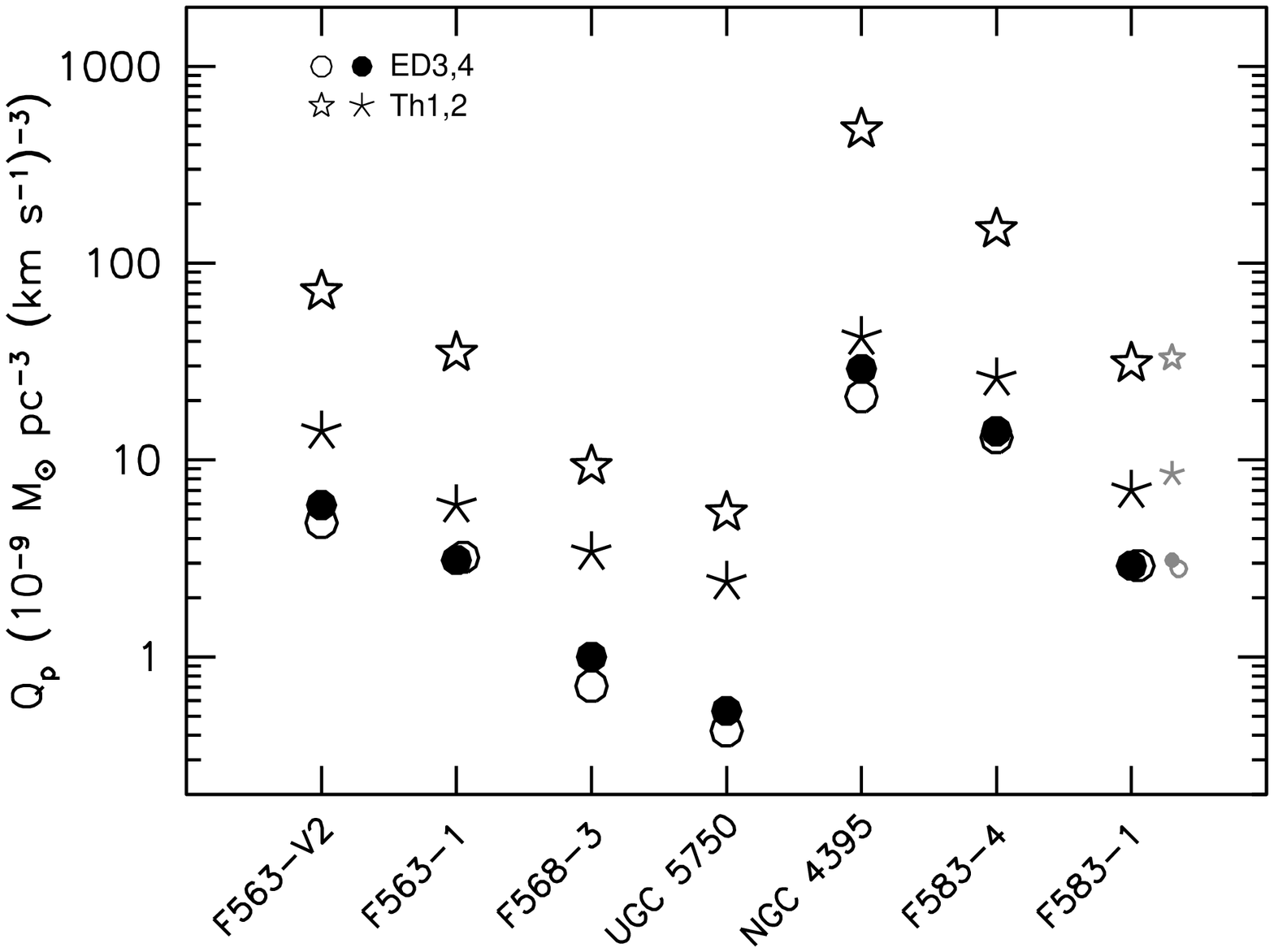}
\caption{The minimum primordial phase space density, $\qp$, inferred
  from each galaxy in the sample.  Each symbol represents a different
  model for the dark matter halo density profile.  For a given model,
  these galaxy data are not consistent with a single value of $\qp$.
  The small gray symbols for F583-1 indicate the results when a
  non-zero stellar mass-to-light ratio is assumed.}
\end{figure}

\section{Cored Dark Matter Halo Fits}

We fit the cored dark matter models to the observed rotation curves of
a sample of nine LSB galaxies.  The rotation curves are derived from
high-resolution H$\alpha$ integral field unit spectroscopy
\citep{Kuzio06,Kuzio08} and are combined with rotation curves derived
from long-slit H$\alpha$ spectra \citep{dBMR,dBB} and H{\sc i}
velocity fields \citep{DMV,Stil,Swatersthesis}.  The inner part of the
rotation curves set the central core density, $\rhocore$, for each
halo model, and the data at large radii fix $\rhocore\rcore^2$.
Because the dark matter is the dominant mass component in these
galaxies at all radii, we neglect the (minimal) contribution to the
observed rotation from the baryons.  To demonstrate that the effect of
the baryons on the fit parameters is indeed small, we include in the
figures a representative example (F583-1) where a stellar
mass-to-light ratio based on galaxy color is assumed \citep{Kuzio08}.

In Figure 1 we plot the four cored halo fits over the data.  The
central density and core radius derived for each dark matter model are
listed in Table 1.

We find that for most galaxies, the best-fitting halo rotation curves
match the observed galaxy rotation curves well.  In general, the early
decay and thermal WDM models produce very similar fits to the data and
there is little dependence on the value of the exponent $\alpha$.
With the exception of the thermal WDM model with $\alpha=2$ for some
of the galaxies, the best-fitting halo rotation curves typically
overlap at most radii.  For the four halo models, the sizes of the
derived cores are on the order of a kpc and span a range of about
a factor of 10.

The observed rotation curves of UGC~4325 and DDO~64 are not
well-described by the halo models ($\chi^{2}_{r} \gtrsim 3$).  Rather
than turning over, the models continue to rise past the last observed
rotation curve point. The implied cores would be larger than the
observed radial range of the data, and the masses of the systems would
be about $10^{12} M_\odot$ or larger. Given the poor fits, we exclude
UGC~4325 and DDO~64 from the remaining analysis.  We note that the
$\chi^{2}_{r}$ values for the NGC~4395 halo fits are also large (2.6 -
5.1). These high values are the result of the features seen in the
rotation curve around 2.5 kpc that cannot be fit by smooth dark matter
density profiles. However, the velocity profiles do follow the overall
shape and turnover of the observed rotation curve.

\section{Discussion}

Given a candidate dark matter particle with primordial phase space
density $\qp$, any halo with total gravitationally bound mass
$\mtot$ must have a minimum constant density core radius that scales
inversely with the total halo mass as $\rcore \propto \qp^{-2/3} \,
\mtot^{-1/3}$ \citep{kaplinghat05,Martinez}. This
relation follows from dimensional analysis given our assumption that
the shape of the density profile only depends on a scale density and
scale radius.  It can also be explicitly calculated in the context of
the ``excess mass function'' of \citet{dehnen05}, as shown by
\citet{kaplinghat05}. Given the total mass and primordial phase
space density, a lower bound on the core size can be placed because
entropy increases or, equivalently, because ``phase space density''
decreases. More correctly, the values of the minimum core size reflect
a situation where the particles have not phase-space mixed.

The lower limit to the primordial average phase space
density for thermal WDM models may be written as
\begin{equation}
\label{eq:thQ}
\qp \gtrsim \left(\matrix{84\cr 71}\right) 
\qunit 
\rhocore^{-1/2} \rcore^{-3},
\;\left[\matrix{\textrm{Th1}\cr \textrm{Th2}}\right]
\end{equation}
where the upper and lower numbers are for Th1 and Th2 models,
respectively, and $\rho_{0}$ is in $\textrm{M}_\odot \textrm{pc}^{-3}$
and $r_{core}$ is in pc.  Note that the dependence on the
actual shape of the halo profile is mild.
 
For models of non-thermal WDM from early decays, the lower limit to
the primordial average phase space density may be written in terms of
the total mass of the dark matter halo as follows 
\begin{equation}
\qp \gtrsim 82 \,\qunit
\sqrt{\frac{M_{\sun}}{\mtot}\left(\frac{pc}{\rcore}\right)^{3}}. \; [\textrm{ED}]
\label{eq:edQ}
\end{equation}

We calculate the lower limits on $\qp$ for each galaxy and each dark
matter model using Equations 5 and 6 and list them in Table 1.  For
the ED4 and Th2 models we also list the total mass of the galaxy-halo
system, $\mtot$, inferred from the best-fit density profile.  The
masses determined in this way are consistent with and only marginally
larger than the enclosed mass determined using the last observed
rotation curve point.  The ED3 and Th1 models formally have divergent
masses because of the assumed profile shape at large r and are not
listed in Table 1.

The halo parameters presented in Table 1 do not seem to obey the
$\rcore \propto \mtot^{-1/3}$ scaling relation that is expected if the
cores are set by the requirement of no phase-space mixing.  The
observed core radii of our galaxy sample span a range of about a
factor of ten, which would require a factor of $\sim 1000$ variation
in total mass from system to system to be explained by the same
primordial phase space density.  Given that these galaxies are so
similar in luminosity and asymptotic rotation speeds (Figure 1), this
is highly implausible.   

We note that this argument does not rule out the possibility that the
underlying dark matter is warm, but only that core sizes are not set
by the primordial phase space density of dark matter particles. In
particular, some of the cores could be due to phase-space mixing as a
result of mergers, but this makes the attribution of cores to
fundamental dark matter physics ambiguous. In addition, we will show
that the values of the primordial phase space density required are
much too small to be consistent with constraints arising from the
matter power spectrum.

In Figure 2, we present a more detailed comparison of theory and data
by plotting the inferred total mass of each galaxy-halo system,
$\mtot$, against the measured halo core radii.  For clarity we plot
only the results for the Early Decay $\alpha=4$ and Thermal WDM
$\alpha=2$ cases. Qualitatively similar results are obtained for the
ED3 and Th1 cases if we use $M_{\rm vir}$ rather than $\mtot$ and do
not modify the conclusions.  For each galaxy, we plot the combined
range of core radii and masses for the two dark matter models.  For
comparison, we have also plotted the $\mtot$ vs. (minimum) $\rcore$
relationship that is expected for early-decay dark matter for two
choices of $\qp$.  It is immediately obvious from Figure 2 that 1) the
data span a range of only about one order of magnitude in mass, 2) the
data are not consistent with a single value of $\qp$, and 3) mass and
core radius are not anti-correlated as would be expected from Equation
\ref{eq:edQ}.

\begin{figure}
\epsscale{1.0}
\plotone{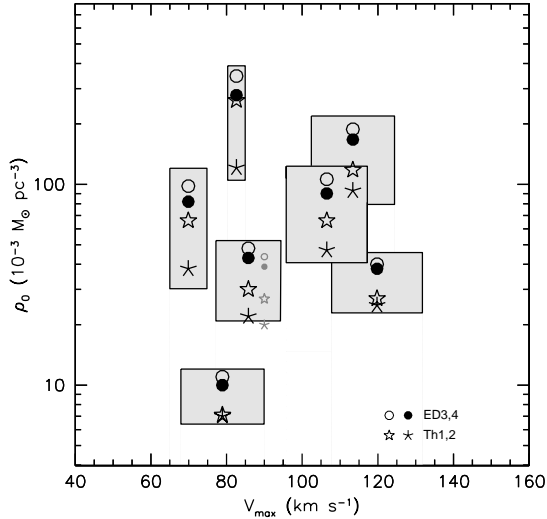}
\caption{Halo central density, $\rhocore$, as a function of the
maximum observed rotation velocity of the galaxy. Each symbol
represents a different model for the dark matter halo density profile.
For a given model, $\rhocore$ is not constant across the sample, and
there is no discernible trend in $\rhocore$ with $\vmax$.  The small
gray symbols indicate the results when a non-zero stellar
mass-to-light ratio is assumed.}
\end{figure}

The simplest interpretation of this result in the context of dark
matter models is that the cores in these galaxies $\textit{cannot}$ be
set directly by the primordial phase space density of dark matter and
therefore must be the result of baryonic processes.  If, however, we
insist that a WDM model explain these data, then to have a single
value of $\qp$ for this sample, galaxies with small cores must
preferentially lose more than 2 orders of magnitude in mass, while
galaxies with large cores lose very little.  This is highly unlikely
in these undisturbed disk galaxies, as feedback from powerful radio
sources is observed to occur almost always only in elliptical galaxies
or obvious recent mergers \citep{Wilson95, Urry95,Antonucci93}.
Additionally, feedback from supernova winds is also unlikely to affect
these galaxies, as the star formation rates in LSBs are known to be
lower than the rates in high surface brightness galaxies of similar
morphological type \citep{Bothun97,Oneil07}.  We note here that recent
high resolution hydrodynamical simulations have produced galaxies with
cored CDM halos by including baryonic processes that effectively
remove mass \citep{Governato09,Mashchenko08}, though \citet{Ceverino}
reach a different conclusion.  Finally, even if there were a plausible
model to explain Figure 2, we show below that the required value of
$\qp$ is in strong disagreement with Lyman alpha forest data.

In Figure 3, we plot the range of $\qp$ for the galaxies and again
find that, for a given dark matter model, the data are \textit{not}
consistent with a single $\qp$ value.  For our sample of galaxies,
$\qp$ ranges between $\sim\,10^{-9}$ and $10^{-7}$ in units of
$\qunit$.  This result does not change when the baryons are accounted
for by assuming a non-zero stellar mass-to-light ratio, as shown for
F583-1 in Figure 3.  These limits on $\qp$ are about 4 orders of
magnitude smaller than the lower limit on thermal WDM implied by the
Lyman alpha forest power spectrum of $\simeq 10^{-3}$ $\qunit$
\citep{Seljak06,Viel08}. Thus, even if there were a WDM model whose
primordial phase space density value was in tandem with some other
process that sets the core sizes in these galaxies, we would have a
model that is inconsistent with the Lyman alpha forest data by orders
of magnitude.

We now consider the SIDM model predictions. This is easier to analyze
because the SIDM models predict a correlation between core size and
core density. In most models of dark matter with large
self-interactions, all dark matter halos are predicted to either have
the same core density or to show a trend in $\rhocore$ as a function
of velocity dispersion of the halo
\citep{Spergel00,Firmani00,Kaplinghat00}. One reason for this trend is
the dependence of the scattering or annihilation cross section on
relative velocity. Additionally, adiabatic expansion due to particle
loss will result in a systematically smaller core density in less
massive halos \citep{Kaplinghat00}. 

We note that a monotonic relation between the cross section and the
velocity translates to a monotonic relation between the core density
and the velocity dispersion of the dark matter particles in the
core. We expect the velocity dispersion in the core to be isotropic
and proportional to $\vmax$. It therefore follows that
the expectation from SIDM models is that the inferred core density
should be either roughly constant or exhibit a monotonic trend with
$\vmax$.  We note that if the self-interaction process has been
operating for differing times in these galaxies, for example as the
result of a recent major merger, then some dispersion may be
introduced into the inferred $\rhocore$ versus $\vmax$ relation.
However, this seem unlikely given the uniformity of the sample and the
lack of observational evidence for any recent disturbance.

In Figure 4, we plot $\rhocore$ against $\vmax$ for each galaxy and
show a representative example of how $\rhocore$ changes if a non-zero
stellar mass-to-light ratio is assumed.  We find that $\rhocore$ is
not constant across the sample, nor is there evidence for a systematic
trend in $\rhocore$ as a function of $\vmax$.  This indicates that the
inferred cores in these LSB galaxies cannot be directly set by large
self-interactions (scattering or annihilation) of dark matter.

\section{Summary}

Warm dark matter models and strongly self-interacting dark matter
models have been proposed to alleviate some of the difficulties that
CDM faces on small scales.  We have tested models of thermal WDM
and non-thermal WDM from early decays with high-resolution rotation
curves for LSB galaxies. We infer the observed halo core radii to span
about an order of magnitude around a kpc, while WDM models would
predict a spread of only about a factor of 2. Additionally, the values
of $\qp$ inferred from these LSB galaxies are orders of magnitude
smaller than the lower limits implied by the Lyman alpha forest power 
spectra. Taken together, we interpret these results to mean that the
cores in these LSB disk galaxies cannot be a direct result of WDM
particle properties.  We also find the data to be inconsistent with a
single value of core density $\rhocore$ and find no evidence for a
trend in $\rhocore$ with the maximum circular velocity.  This strongly
argues against the possibility that large self-interactions of dark
matter are directly responsible for setting the cores.

\acknowledgements  R.K.D. is supported by an NSF Astronomy \&
Astrophysics Postdoctoral Fellowship under award AST 07-02496.
R.K.D. thanks Louie Strigari for helpful early discussions. M.K.,
J.S.B., and G.D.M. are supported in part by NSF AST-0607746 and NASA
grant NNX09AD09G.

\end{document}